\documentclass[11pt,a4paper]{article}
\usepackage{a4,amsmath,cite}

\begin{document}

\title{A new approach to the relativistic treatment
of the fermion-boson system, based on the extension of the
$SL(2,C)$ group}

\author{D.A. Kulikov\thanks{kulikov@ff.dsu.dp.ua},
R.S. Tutik\thanks{tutik@ff.dsu.dp.ua (corresponding author)}\\
\\
Theoretical Physics Department \\ Dniepropetrovsk National
University \\
72 Gagarina av., Dniepropetrovsk 49050, Ukraine }
\date{}

\maketitle

\begin{abstract}
A new technique for constructing the relativistic wave equation
for the two-body system composed of the spin-1/2 and spin-0
particles is proposed. The method is based on the extension of the
$SL(2,C)$ group to the $Sp(4,C)$ one. The obtained equation
includes the interaction potentials, having both the
Lorentz-vector and Lorentz-tensor structure, exactly describes the
relativistic kinematics and possesses the correct one-particle
limits. The comparison with results of other approaches to this
problem is discussed.
\end{abstract}

\section{\protect Introduction}
\label{part1}

Necessity of taking into account the relativistic effects in
two-body systems arises in many problems in particle and nuclear
physics. In recent years, there has been much interest in the
relativistic treatment of the two-body problem concerning the
spin-1/2 fermion and the spin-0 boson. For such systems, finite
size effects and relativistic corrections in the pionic and kaonic
hydrogen have been evaluated \cite{kelkar}, effects of the
retarded interaction between an electron and a spin-0 nucleus have
been studied \cite{datta}. These investigations apply the
technique of the Breit equation \cite{breit}, so that expansions
in powers of $1 /c^2$ are used. Alternatively, there exist other
approaches to the fermion-boson problem that describe the
relativistic kinematics exactly. Namely, the phenomenological
two-body equation proposed by Kr\'olikowski \cite{krolikowski},
the Barut method \cite{barut,klimek}, the reductions of the
Bethe-Salpeter equation \cite{tanaka,bijtebier96} and the
relativistic quantum mechanics with constraints
\cite{sazdjian86a,sazdjian86b,crater87,crater04} are known.

Note that the above-mentioned investigations deal with potentials
of the Lorentz-vector structure. However, the tensor coupling
receives considerable attention of late. For instance, within the
framework of the one-particle Dirac equation, the Lorentz-tensor
potentials have been used for describing the nuclear properties
\cite{furnstahl,mao,alberto} and for constructing the relativistic
harmonic-oscillator models \cite{moshinsky,kukulin}.

Recently, the of possibility of deriving the relativistic wave
equation for the two-body problem by means of the extension of the
$SL(2,C)$ group to the $Sp(4,C)$ one has been proved
\cite{kulikov}. The goal of the present work is to elaborate the
proposed technique and to construct the relativistic equation for
the two-body system consisted of the spin-$1/2$ and spin-$0$
particles. This equation must be suitable for including not only
the commonly used Lorentz-vector potentials but the Lorentz-tensor
ones, too, that will permit us to go beyond the one-particle
approximation in the consideration of the tensor coupling effects
in the fermion-boson systems.

The paper is organized as follows. Section \ref{part2} is devoted
to a symplectic spacetime extension generated by the extension of
the $SL(2,C)$ group to the $Sp(4,C)$ one. In Section \ref{part3}
this approach is applied to construct the relativistic wave
equation for a fermion-boson system with interaction introduced by
means of the Lorentz-vector and -tensor potentials. The comparison
with results of other approaches is given in Section \ref{part4}.
At last, our conclusions are summarized in Section \ref{part5}.

\section{Symplectic spacetime extension}
\label{part2}

It is known that the homogeneous Lorentz group $SO(1,3)$ is
covered by the $Sp(2,C)\equiv SL(2,C)$ group. As a consequence,
there exists the one-to-one correspondence between the $Sp(2,C)$
Hermitian spin-tensors of the second rank and the Minkowski
four-vectors. Therefore, the relativistic field theory in the
Minkowski space can be equivalently formulated entirely through
the $Sp(2,C)$ Weyl spinors \cite{penrose}.

Developing this approach, we may consider the symplectic
$Sp(2l,C)$, $l>1$ group which is linked to a $(2l)^2$-dimensional
metric spacetime with a signature corresponding to $l(2l-1)$
timelike and $l(2l+1)$ spacelike dimensions \cite{pirogov}. Note
that this extension of the pseudo-Euclidean spacetime is
alternative to those having one time and $d>3$ spatial dimensions
and used in the Kaluza-Klein theory. For description of the
two-particle systems, the suitable symplectic spacetime extension
has been proved to be the extension with the $Sp(4,C)$ group
\cite{kulikov}. For completeness and consistency of the following
consideration, let us now outline the main ideas of this approach.

The symplectic $Sp(4,C)$ group is the group of $4\times 4$
matrices, with complex elements and determinant equal to one,
acting on the four-component Weyl spinors $\varphi_{\alpha}$ and
preserving an antisymmetric bilinear form
$\eta_{\alpha\beta}=-\eta_{\beta\alpha}$ \cite{weyl}. This form
plays the role of the ``metrics" in the spinor space in the sense
that the Weyl spinors with lower and upper indices are related by
transformation
$\varphi_{\alpha}=\eta_{\alpha\beta}\varphi^{\beta}$. This means
that the representations $\varphi_{\alpha}$ and $\varphi^{\alpha}$
are equivalent, but the complex conjugative spinors
$\bar{\varphi}_{\bar{\alpha}}=(\varphi_{\alpha})^*$ belong to the
other representation. Thus, we have two spinor representations,
say $\varphi_{\alpha}$ and $\bar{\chi}^{\bar{\alpha}}$, which are
suitable for describing the wave functions.

In order to introduce the spacetime, in which the wave functions
will be defined, we consider the irreducible representation of the
$Sp(4,C)$ group described by Hermitian spin-tensor of the second
rank, $P_{\alpha\bar{\alpha}}$. As in the case of the $Sp(2,C)$
group, such spin-tensors also correspond to real vectors but
having already 16 components. Let us define this correspondence by
the relations
\begin{equation}\label{eq1}
P_{\alpha\bar{\alpha}}=\mu^M_{\alpha\bar{\alpha}}P_M, \qquad
P^M=\frac{1}{4}\tilde{\mu}^{M\bar{\alpha}\alpha}P_{\alpha\bar{\alpha}}
\end{equation}
where $\mu^M_{\alpha\bar{\alpha}}$ ($M=1,2,...,16$) are matrices
of the basis in the space of $4 \times 4$ Hermitian matrices and
tilde labels the transposed matrix with upper spinor indices.
Hereafter we will suppress the spinor indices when possible.

For clarifying the connection of the discussed vector space
$\mathsf{R^{16}}$ to the Min\-kow\-s\-ki space $\mathsf{R^{4}}$,
let us represent 16 values of the vector index of $P_M$ through
$4\times 4$ combinations of two indices, $M=(a,m)$, with both $a$
and $m$ running from $0$ to $3$. Then the matrices $\mu^M$ and
$\tilde{\mu}^N$ can be chosen as follows
\begin{equation}\label{eq2}
\mu^M\equiv\mu^{(a,m)}=\Sigma^a\otimes\sigma^m, \qquad
\tilde{\mu}^N\equiv\tilde{\mu}^{(b,n)}=\tilde{\Sigma}^b\otimes\tilde{\sigma}^n,
\end{equation}
where explicit expressions for matrices $\Sigma^a$ and $\sigma^m$,
written in terms of $2\times 2$ unit matrix $I$ and the Pauli
matrices $\tau^i$, are
\begin{eqnarray}\label{eq2a}
\Sigma^0=\tilde{\Sigma}^0=I, \quad
\Sigma^1=\tilde{\Sigma}^1=\tau^1, \quad
\Sigma^2=-\tilde{\Sigma}^2=\tau^2, \quad \Sigma^3=\tilde{\Sigma}^3=\tau^3, \nonumber \\
\sigma^0=\tilde{\sigma}^0=I, \quad
\sigma^1=-\tilde{\sigma}^1=\tau^1, \quad
\sigma^2=-\tilde{\sigma}^2=\tau^2, \quad
\sigma^3=-\tilde{\sigma}^3=\tau^3.
\end{eqnarray}

With such the representation for matrices $\mu^M$ and
$\tilde{\mu}^N$, the metrics
$g^{MN}=\frac{1}{4}\mu^M_{\alpha\bar{\alpha}}\tilde{\mu}^{N\bar{\alpha}\alpha}$
of the discussed vector space $\mathsf{R^{16}}$ explicitly takes
the factorized form
\begin{equation}\label{eq3}
g^{MN}\equiv g^{(a,m)(b,n)}=h^{ab}h^{mn}
\end{equation}
where $h^{mn}=diag(1,-1,-1,-1)$ is the usual Minkowski metrics and
the factor $h^{ab}=diag(1,1,-1,1)$ is caused by the group
extension.

The factorization of the metrics implies that a vector from
$\mathsf{R^{16}}$ may be decomposed into four Minkowski
four-vectors. As a consequence, the $Sp(4,C)$ momentum spin-tensor
(\ref{eq1}) can be employed to construct the wave equation for a
two-body system.

\section{Wave equation for a fermion-boson system}
\label{part3}

We are going to construct the wave equation for the system
consisted of one spin-1/2 and one spin-0 particle. With the total
spin of the system being equal to $1/2$, the wave function of this
system have to be represented by a Dirac spinor or, alternatively,
by two Weyl spinors \cite{penrose}. Then, in terms of the
$Sp(4,C)$ Weyl spinors, the corresponding wave equation must have
the form of the Dirac equation
\begin{equation}\label{eq4}
P_{\alpha\bar{\alpha}}\bar{\chi}^{\bar{\alpha}}=m\varphi_{\alpha},
\qquad
\tilde{P}^{\bar{\alpha}\alpha}\varphi_{\alpha}=m\bar{\chi}^{\bar{\alpha}}
\end{equation}
where $m$ is a mass parameter and $P_{\alpha\bar{\alpha}}$ is the
$Sp(4,C)$ momentum spin-tensor which, in view of Eqs.(\ref{eq1})
and (\ref{eq2}), can be written as
\begin{equation}\label{eq5}
P=\mu^{(a,m)}P_{(a,m)}=\Sigma^0\otimes\sigma^m
w_m+\Sigma^1\otimes\sigma^m p_m+ \Sigma^2\otimes\sigma^m
u_m+\Sigma^3\otimes\sigma^m q_m
\end{equation}
where $w_m$, $p_m$, $u_m$, $q_m$ are the Minkowski four-momenta.
As it has been shown \cite{kulikov}, this wave equation,
supplemented with subsidiary conditions, describes the
fermion-boson system with the equal mass constituents.

Now we generalize the wave equation (\ref{eq4}) to the case of
particles with unequal masses. For this end, let us replace the
mass parameter on the right-hand side of this equation by a
suitable matrix term which breaks the $Sp(4,C)$ symmetry of the
wave equation but retains the Lorentz $SO(1,3) \subset Sp(4,C)$
symmetry. A natural ansatz for this term is a combination of
direct products of matrices as in Eq.~(\ref{eq5}). Then the
Lorentz symmetry is retained if and only if the second matrix in
this direct product is the unit matrix. In this case, we have two
equivalent possibilities to obtain the plus sign for the quantity
$h^{ab}$, which are realized by choosing the first matrix as
$\tau^1$ or $\tau^3$. In view of this fact, we replace the mass
parameter as follows:
\begin{equation}\label{eq100}
m\rightarrow(m_1+m_2)/2+\tau^1\otimes I (m_1-m_2)/2,
\end{equation}
so that the additional term vanishes if $m_1=m_2$.

Thus, the wave equation without interaction for the fermion-boson
system with unequal masses takes the form
\begin{equation}\label{eq200}
P\bar{\chi}=(m_{+}+\tau^1\otimes I m_{-})\varphi, \qquad
\tilde{P}\varphi=(m_{+}+\tau^1\otimes I m_{-})\bar{\chi}
\end{equation}
where $m_{\pm}=(m_{1}\pm m_{2})/2$.

To proceed, let us consider the structure of the $Sp(4,C)$
momentum spin-tensor given by Eq.~(\ref{eq5}). It should be
stressed that the description of the two-particle system requires
only two four-momenta whereas the $Sp(4,C)$ momentum spin-tensor
involves four four-momenta, $w_m$, $p_m$, $u_m$, and $q_m$. Hence,
the number of their independent components must be decreased that
can be implemented by means of subsidiary conditions.

For deriving the subsidiary conditions, we transform
Eqs.~(\ref{eq200}) into the form of the Klein-Gordon equation that
guarantees the correct dynamical relation between energy and
momentum for free particles. Upon eliminating $\bar{\chi}$ and
inserting Eq.~(\ref{eq5}), we arrive at
\begin{equation}\label{eq6}
\left(w^2+p^2-u^2+q^2-\frac{2m_{-}}{m_{+}}(wp-iuq)-m_{+}^2+m_{-}^2+\sum^{5}_{A=1}\Gamma_A
K^A\right)\varphi=0
\end{equation}
where $w^2=(w^0)^2-\mathbf{w}^2$, $wp=w^0
p^0-\mathbf{w}\mathbf{p}$, etc, $\Gamma_A$ are direct products of
the Pauli matrices, and $K^A$ are quadratic forms with respect to
the four-momenta.

Because the non-diagonal terms $\Gamma_A K^A$ in this equation do
not occur in the case of the ordinary Klein-Gordon equation, we
put $K^A=0$ on the wave functions $\varphi$ and $\bar{\chi}$ that
yields
\begin{eqnarray}\label{eq12}
&&[(m_{+}^{2}-m_{-}^{2})(wp-m_{+}m_{-})-m_{+}m_{-}(u^2-q^2)]\psi=0, \nonumber \\
&&(m_{+}wq-m_{-}pq)\psi=0, \nonumber \\
&&(m_{+}up-m_{-}uw)\psi=0,  \\
&&uq\psi=0, \nonumber \\
&&[m_{+}(u^m w^n - u^n w^m -\epsilon^{mnkl}p_k q_l)-m_{-}(u^m p^n
- u^n p^m -\epsilon^{mnkl}w_k q_l)]\psi=0 \nonumber
\end{eqnarray}
where, for brevity, we denote $\psi= \left(\begin{array}{c}
   \varphi \\
    \bar{\chi}
  \end{array}\right)$ and $\epsilon^{mnkl}$ is the totally antisymmetric
tensor ($\epsilon^{0123}=+1$).

Thus, the imposed conditions and the Klein-Gordon-like equation
(\ref{eq6}) set ten components of $w_m$, $p_m$, $u_m$, $q_m$ to be
the independent ones. The connection of these four-momenta with
the four-momenta, $p_{1m}$ and $p_{2m}$, of the constituent
particles is established by assuming
\begin{equation}\label{eq300}
w_m=\frac{1}{2}(p_{1m}+p_{2m}),\quad
p_m=\frac{1}{2}(p_{1m}-p_{2m}),\quad u_m=0,\quad q_m=0.
\end{equation}
With these expressions for the four-momenta, the only one
condition in Eqs.~(\ref{eq12}) remains nontrivial that reads
\begin{equation}\label{eq13}
(wp-m_{+}m_{-})\psi=0.
\end{equation}

This constraint, on the one hand, together with the
Klein-Gordon-like equation guarantees that for the free particle
case, since $wp-m_{+}m_{-}=(p_1^2-p_2^2-m_1^2+m_2^2)/4$, the
particles are on the mass shell. But on the other hand it removes
the dependence of the wave function on the relative time.

Note that for the isolated system the total four-momentum $w_m$ is
conserved and so can be treated as the eigenvalue rather than the
operator. As the relative four-momentum can be represented by the
operator $p_m=i\partial/\partial x^m$, with $x^m=x_{1}^m-x_{2}^m$
being the relative coordinate divided in the longitudinal and
transverse parts
\begin{equation}\label{eq400}
x_{\bot}^m=(h^{mn}-w^m w^n/w^2)x_n, \qquad
x_{||}^m=x^m-x_{\bot}^m,
\end{equation}
the subsidiary condition (\ref{eq13}) determines the dependence of
the wave function on the relative coordinates in the form
\begin{equation}\label{eq500}
\psi(x)=e^{-im_{+}m_{-}(wx_{||})/w^2}\psi(x_{\bot}).
\end{equation}
Here the longitudinal part $x_{||}^m$ plays the role of the
``relative time" of the particles, because in the center-of-mass
frame it reduces to the difference of the usual time components of
$x_{1}^m$ and $x_{2}^m$, whereas the transverse coordinates
$x_{\bot}^m$ describe the interparticle separation in space.
However, the relative-time variable is contained only in the phase
factor which may be suppressed. Thus, the dynamics of the relative
motion of this system is described only with the transverse
coordinates $x_{\bot}^m$ and does not depend on the relative time
$x_{||}^m$ of the particles. This property also holds in the case
with the potential interaction included.

In order to clarify the two-particle interpretation of the wave
equation (\ref{eq200}), let us reduce it to the one-particle Dirac
and Klein-Gordon equations which must describe the constituents of
this system. With decomposing the spinor wave functions into the
projections
\begin{equation}\label{eq14}
\varphi_{\pm}=\frac{1}{2}(1\pm\tau^1\otimes I)\varphi, \qquad
\bar{\chi}_{\pm}=\frac{1}{2}(1\pm\tau^1\otimes I)\bar{\chi}
\end{equation}
which are two-component $Sp(2,C)$ Weyl spinors, Eqs.(\ref{eq200})
and (\ref{eq13}) are reduced to two uncoupled sets of equations:
\begin{equation}\label{eq15}
p_{1m}\sigma^m\bar{\chi}_{+}=m_1\varphi_{+}, \qquad
p_{1m}\tilde{\sigma}^m\varphi_{+}=m_1\bar{\chi}_{+}
\end{equation}
\begin{equation}\label{eq15a}
(p_2^2-m_2^2)\varphi_{+}=0, \qquad (p_2^2-m_2^2)\bar{\chi}_{+}=0
\end{equation}
and
\begin{equation}\label{eq16}
p_{2m}\sigma^m\bar{\chi}_{-}=m_2\varphi_{-}, \qquad
p_{2m}\tilde{\sigma}^m\varphi_{-}=m_2\bar{\chi}_{-},
\end{equation}\begin{equation}\label{eq16a}
(p_1^2-m_1^2)\varphi_{-}=0, \qquad (p_1^2-m_1^2)\bar{\chi}_{-}=0.
\end{equation}
These equations contain the free one-particle Dirac equations
written in the Weyl spinor formalism and the free Klein-Gordon
ones. Consequently, the wave equation (\ref{eq200}) supplemented
with the subsidiary condition (\ref{eq13}) describes two systems
composed of the spin-$1/2$ and spin-$0$ particles, which differ
from each other only in the permutation of masses of the
particles.

As a next step, we must incorporate a potential interaction in the
consideration. A widely accepted receipt for introducing the
interaction consists in the replacement of the constituent
four-momenta in the equation without interaction by the
generalized momenta in the minimal manner, so that each particle
is in the presence of a field of an external potential raised by
the other particle. However, there also exists the non-minimal
scheme \cite{moshinsky,kukulin} which uses the momentum
substitution having the matrix structure.

Because the free particle case in our approach is described by the
the equation (\ref{eq200}) and the subsidiary condition
(\ref{eq13}), we introduce the interaction in those through the
combination of the minimal and non-minimal substitutions on the
four-momenta. In terms of the total and relative four-momenta,
$w_m$ and $p_m$, this reads as
\begin{equation}\label{eq17}
\begin{split}
w_m\varphi &\rightarrow
(w_m+A_m)\varphi +iC_{mn}I\otimes\sigma^n\bar{\chi}, \\
w_m\bar{\chi} &\rightarrow
(w_m+A_m)\bar{\chi}+iC_{mn}I\otimes\tilde{\sigma}^n\varphi, \\
p_m\varphi &\rightarrow
(p_m+B_m)\varphi +iD_{mn}I\otimes\sigma^n\bar{\chi}, \\
p_m\bar{\chi} &\rightarrow
(p_m+B_m)\bar{\chi}+iD_{mn}I\otimes\tilde{\sigma}^n\varphi
\end{split}
\end{equation}
where the involved potentials $A_m$, $B_m$, $C_{mn}$ and $D_{mn}$
in the general case are the functions of $w_m$, $p_m$ and the
relative coordinate $x_m$. The non-minimal part of these
substitutions is implemented through the matrices $\sigma^n$ and
$\tilde{\sigma}^n$ which are the blocks of the Dirac
gamma-matrices in the Weyl spinor formalism.

With Eqs.(\ref{eq17}) inserted, the wave equation (\ref{eq200})
takes the form
\begin{equation}\label{eq18}
\begin{split}
[I\otimes \sigma^m (w_m&+A_m)+\tau^1\otimes \sigma^m
(p_m+B_m)]\bar{\chi}                    \\
&=(m_{+}+\tau^1\otimes I m_{-} -iI\otimes\sigma^m\tilde{\sigma}^n
C_{mn}
-i\tau^1\otimes\sigma^m\tilde{\sigma}^n D_{mn})\varphi,  \\
[I\otimes \tilde{\sigma}^m
(w_m&+A_m)+\tau^1\otimes\tilde{\sigma}^m
(p_m+B_m)]\varphi  \\
&=(m_{+}+\tau^1\otimes I m_{-} -iI\otimes\tilde{\sigma}^m\sigma^n
C_{mn} -i\tau^1\otimes\tilde{\sigma}^m\sigma^n D_{mn})\bar{\chi}.
\end{split}
\end{equation}

For elucidating the meaning of the Lorentz-tensor interaction, let
us consider the matrix structure of the terms
$iC_{mn}I\otimes\tilde{\sigma}^m\sigma^n$ and
$iD_{mn}I\otimes\sigma^m\tilde{\sigma}^n$. The algebra of the
matrices $\sigma^m$, $\tilde{\sigma}^m$
\begin{equation}\label{eq20}
\sigma^m\tilde{\sigma}^n=h^{mn}I+\frac{i}{2}\epsilon^{mnkl}\sigma_k\tilde{\sigma}_l,
\qquad
\tilde{\sigma}^m\sigma^n=h^{mn}I-\frac{i}{2}\epsilon^{mnkl}\tilde{\sigma}_k\sigma_l
\end{equation}
implies that the symmetric parts of $C_{mn}$ and $D_{mn}$, being
contracted with the Minkowski metrics $h^{mn}$, result in the
known Lorentz-scalar interaction. In one's turn, the contribution
of the antisymmetric parts of $C_{mn}$ and $D_{mn}$ is similar to
that of the non-minimal coupling term introduced by Pauli
\cite{pauli} within the framework of the one-particle Dirac
equation. As the Pauli term describes the coupling of the
anomalous magnetic moment of a fermion with an external field, the
antisymmetric Lorentz-tensor potentials $C_{mn}$ and $D_{mn}$ can
be regarded as being responsible for the interactions with the
anomalous magnetic moment in the fermion-boson system.

As regards the subsidiary condition, upon inserting (\ref{eq17})
into (\ref{eq13}) and symmetrizing the resulted expression for
each its piece to become self-adjoint, we obtain
\begin{eqnarray}\label{eq181}
&[\frac{1}{2}(\omega_m\pi^m+\pi^m\omega_m)+\frac{i}{2}(\omega^m
D_{mn}-D_{mn}\omega^m+ C_{mn}\pi^m - \pi^mC_{mn})\gamma^n \nonumber \\
&-\frac{1}{2}(C_{mk}D^{mn}+D_{mk}C^{mn})\gamma^k\gamma_n-m_{+}m_{-}]\psi=0
\end{eqnarray}
where the following abbreviations are used
\begin{equation}\label{eq182}
\omega_m=w_m+A_m, \qquad \pi_m=p_m+B_m, \qquad \gamma^m=
\left(\begin{array}{cc}
   0 & \sigma^m\\
   \tilde{\sigma}^m &0
  \end{array} \right)
\end{equation}
and the gamma-matrices are self-adjoint in the usual Dirac sense,
that is
$\bar{\gamma}^m\equiv\gamma^0(\gamma^m)^{\dag}\gamma^0=\gamma^m$.

It should be stressed that the subsidiary condition (\ref{eq181})
must be compatible with the wave equation (\ref{eq18}). Let us
first consider the case when only the Lorentz-vector potentials
are present. Then Eq.~(\ref{eq181}) becomes
\begin{equation}\label{eq183}
[\frac{1}{2}(\omega\pi+\pi\omega)-m_{+}m_{-}]\psi=0
\end{equation}
and the compatibility takes place when the operator
$(\omega\pi+\pi\omega)$ in the subsidiary condition commutes with
all operators in the wave equation:
\begin{equation}\label{eq184}
[\omega\pi+\pi\omega,w_m+A_m]=0, \qquad
[\omega\pi+\pi\omega,p_m+B_m]=0.
\end{equation}
The last conditions, because $[wp,x_m]\neq 0$ but $[wp,x_{\bot
m}]=0$, require that the potentials $A_m$ and $B_m$ depend on the
relative coordinate only through its transverse part $x_{\bot m}$
defined by Eq.~(\ref{eq400}). Then the simplest solution to the
compatibility conditions (\ref{eq184}) comes from the ansatz
\begin{equation}\label{eq185}
\omega\pi+\pi\omega=2wp
\end{equation}
that at once results in vanishing commutators (\ref{eq184}), and
the subsidiary condition (\ref{eq183}) takes the same form as
Eq.(\ref{eq13}) describing the case without interaction.
Therefore, in the presence of the interaction, the dependence of
the wave function on the relative-time variable can be excluded in
the same manner as in the case without interaction.

The ansatz (\ref{eq185}) obviously restricts the shape of the
potentials $A_m$ and $B_m$. As it will be shown, with the
potentials so involved, the interaction is described both by the
time-component of the Lorentz vector and its spacelike part, as
proposed in the two-particle relativistic quantum mechanics with
constraints \cite{sazdjian86a,sazdjian86b,crater87}. However, the
wave equation (\ref{eq18}) enables us to treat not only the
commonly used Lorentz-vector potentials but also the novel
interaction, involved through the Lorentz-tensor potentials
$C_{mn}$ and $D_{mn}$.

In the case of the Lorentz-tensor interaction, from
Eqs.~(\ref{eq181}) and (\ref{eq183}) it follows that the shape of
the tensor potentials is restricted by the relations
\begin{equation}\label{eq186}
\omega^m D_{mn}-D_{mn}\omega^m+ C_{mn}\pi^m - \pi^mC_{mn}=0,
\qquad C_{mk}D^{mn}+D_{mk}C^{mn}=0.
\end{equation}
Then the requirement of commuting the operator
$(\omega\pi+\pi\omega)$ from the subsidiary condition with the
Lorentz-tensor potentials $C_{mn}$ and $D_{mn}$ in the wave
equation is easily satisfied by the demand that the potentials
$C_{mn}$ and $D_{mn}$ depend on the relative coordinate only
through its transverse part $x_{\bot m}$.

Thus, the subsidiary condition, supplementing the wave equation,
has the same form (\ref{eq13}) in both cases, with and without
interaction. It is the condition that enables us to exclude the
dependence of the wave function on the relative-time variable.

However, it is to be pointed that in spite of having the same
form, this condition has the different physical content. So, in
the presence of the interaction the particles are not on the mass
shell because the form of the wave equation is changed.
Furthermore, the relative four-momentum $p_m$ is not conserved
since it does not commute with the potentials of the wave equation
(for instance, $[p_m,A_n]=i\partial A_n/\partial x_{\bot}^m\neq 0$
though $[wp,A_n]=0$).

\section{Comparison with results of other approaches}
\label{part4}

Now let us compare the proposed equation (\ref{eq18}) with those
obtained with some other approaches to this problem.

At first we discuss the connection with the equations derived
within the framework of the relativistic quantum mechanics with
constraints \cite{sazdjian86a,sazdjian86b,crater87}. Because the
wave equation (\ref{eq18}) describes two fermion-boson systems
which differ from each other only in the permutation of masses of
the particles, we define the first particle as a Dirac fermion
with the mass $m_1$ and the second one as a Klein-Gordon boson
with the mass $m_2$. According to the decomposition (\ref{eq14}),
the wave functions of this system are the spinor projections
$\varphi_{+}$ and $\bar{\chi}_{+}$. If we now use the
representation
\begin{equation}\label{eq21}
\Psi= \left(\begin{array}{c}
   \varphi_{+} \\
   \bar{\chi}_{+}
  \end{array} \right), \qquad
\gamma^m= \left(\begin{array}{cc}
   0 & \sigma^m\\
   \tilde{\sigma}^m &0
  \end{array} \right)
\end{equation}
and take into account definitions (\ref{eq300}) our equation
(\ref{eq18}), in the lack of the Lorentz-tensor interaction, can
be readily transformed into the Dirac-like form
\begin{equation}\label{eq22}
(p_{1m}\gamma^m-m_1-V)\Psi=0
\end{equation}
where $V=-(A_m+B_m)\gamma^m$. Together with the subsidiary
condition (\ref{eq13}), this yields
\begin{equation}\label{eq23}
[p_{2}^2-m_2^2-(p_{1m}\gamma^m+m_1)V]\Psi=0.
\end{equation}
These equations are just those obtained for the fermion-boson
system within the framework of the two-particle relativistic
quantum mechanics with constraints \cite{sazdjian86a,sazdjian86b}.

However, in the relativistic quantum mechanics, there is another
approach \cite{crater87} for involving the interaction in the pair
of the Dirac and Klein-Gordon equations that uses the generalized
four-momenta of the particles $\pi_{im}=p_{im}+A_{im}\, (i=1,2)$
instead of the single potential $V$. In such procedure, the shape
of the potentials $A_{1m}$ and $A_{2m}$ is restricted by the
compatibility requirement and the demand to give correct
one-particle limits for these equations.

In our treatment, such method is easily reproduced with the
replacement of the Lorentz-vector potentials $A_m$ and $B_m$,
introduced by the substitution (\ref{eq17}) on the total and
relative four-momenta, by their expressions in terms of $A_{1m}$
and $A_{2m}$ as
\begin{equation}\label{eq24}
A_m=(A_{1m}+A_{2m})/2, \qquad B_m=(A_{1m}-A_{2m})/2.
\end{equation}
Note that the above-mentioned restrictions on the shape of the
potentials $A_{1m}$ and $A_{2m}$ are reduced to the constraint
(\ref{eq185}) in our method.

Then, adopting the ansatz \cite{crater87}
\begin{eqnarray}\label{eq25}
A_m&=&\left(\left(1-\frac{2\mathcal{A}}{E}\right)^{1/2}-1\right)w_m, \nonumber \\
B_m&=&\left(\left(1-\frac{2\mathcal{A}}{E}\right)^{-1/2}-1\right)p_m
+\frac{i}{2E}\left(1-\frac{2\mathcal{A}}{E}\right)^{-3/2}\frac{\partial
\mathcal{A}}{\partial x_{\bot}^m}
\end{eqnarray}
where $E=2\sqrt{w^2}$ is the total energy and
$\mathcal{A}=\mathcal{A}(x_{\bot}^2)$ is a scalar potential
function, and inserting these potentials in our Eqs.~(\ref{eq22})
and (\ref{eq23}) results in the coupled Dirac and Klein-Gordon
equations with the ``electromagneticlike" Lorentz-vector
interaction derived in the framework of the relativistic quantum
mechanics with constraints \cite{crater87}.

Another approach to describing the fermion-boson system consists
in the use of the effective single-body equation in the
center-of-mass frame \cite{krolikowski,tanaka}.

For obtaining such an equation within the framework of our
formalism, we consider the center-of-mass frame with
$\mathbf{w}\equiv(\mathbf{p}_1+\mathbf{p}_2)/2=0$ and $E=2w_0$.
From Eq.~(\ref{eq13}) it follows that the zeroth component of the
relative four-momentum is $p_0=(m_1^2-m_2^2)/2E$. Then, using the
standard notation for the Dirac matrices, $\gamma^0=\beta$,
$\boldsymbol\gamma=\beta\boldsymbol\alpha$, the equation
(\ref{eq22}) with the potentials (\ref{eq25}) is rewritten as
\begin{eqnarray}\label{eq26}
\Bigl[(1-2\mathcal{A}/E)^{-1/2}\boldsymbol\alpha\mathbf{p}+\beta
m_1
-\frac{E^2-2E\mathcal{A}+m_1^2-m_2^2}{2E(1-2\mathcal{A}/E)^{1/2}}
-\frac{i\boldsymbol\alpha\boldsymbol\nabla
\mathcal{A}}{2E(1-2\mathcal{A}/E)^{3/2}}\Bigr]\Psi=0. \nonumber \\
\notag
\end{eqnarray}
\begin{equation}\label{eq26}
\phantom{hello}
\end{equation}
On the other hand, if we take the other form of the potentials
\begin{eqnarray}\label{eq27}
A_0={\displaystyle -\frac{\mathcal{V}}{E}w_0}, \qquad
\mathbf{A}=0, \qquad
B_0={\displaystyle\frac{\mathcal{V}}{E-\mathcal{V}}p_0}, \qquad
\mathbf{B}= \frac{i\boldsymbol\nabla\mathcal{V}}{2(E-\mathcal{V})}
\end{eqnarray}
we arrive at the equation \cite{krolikowski}
\begin{equation}\label{eq28}
\left[\boldsymbol\alpha\mathbf{p}+\beta
m_1-\frac{E-\mathcal{V}}{2}-\frac{m_1^2-m_2^2-i\boldsymbol\alpha\boldsymbol\nabla
\mathcal{V}}{2(E-\mathcal{V})}\right]\Psi=0
\end{equation}
which can also be derived by reducing the Bethe-Salpeter equation
\cite{tanaka}.

It is interesting to study the difference between Eqs.~
(\ref{eq26}) and (\ref{eq28}) caused by the different structure of
the involved potentials. Note that Eq.~(\ref{eq28}) includes only
the timelike Lorentz-vector interaction, whereas Eq.~(\ref{eq26})
involves the potentials $A_m$ and $B_m$ having both the timelike
and spacelike parts, as seen from their manifestly covariant form
(\ref{eq25}). If we set the potential $\mathcal{A}$ in
Eq.~(\ref{eq26}) as $\mathcal{A}=\mathcal{V}-\mathcal{V}^2/2E$,
the above two equations will have the same form, except for the
derivative terms with $\boldsymbol\nabla \mathcal{A}$ and
$\boldsymbol\nabla \mathcal{V}$, and the factor before
$\boldsymbol\alpha\mathbf{p}$. The difference in the derivative
terms is unessential because these terms can be removed by passing
to new wave functions, namely,
$\Psi_1=(1-2\mathcal{A}/E)^{-1/4}\Psi$ in Eq.~(\ref{eq26}). On the
contrary, the difference between Eqs.~(\ref{eq26}) and
(\ref{eq28}) caused by the factor before
$\boldsymbol\alpha\mathbf{p}$ is meaningful and has its origin in
the Lorentz-structure of the potential interaction in these
equations.

However, in spite of this, Eqs.~(\ref{eq26}) and (\ref{eq28})
coincide in either one-particle limits, i.e., when the mass of one
of particles is much larger than the other. In the case of the
heavy spin-0 particle, putting $E=m_2+E_1$ and taking the limit
$m_2/m_1\rightarrow\infty$, Eq.~(\ref{eq26}) as well as
Eq.~(\ref{eq28}) reduces to the one-particle Dirac equation
\begin{equation}\label{eq29}
(\boldsymbol\alpha\mathbf{p}+\beta m_1+\mathcal{V}-E_1)\Psi=0.
\end{equation}

In the other limit, $m_1/m_2\rightarrow\infty$, if we write
$E=m_1+E_2$, both Eqs.~(\ref{eq26}) and (\ref{eq28}) give rise to
the one-particle Klein-Gordon equation
\begin{equation}\label{eq30}
[\mathbf{p}^2+m_2^2-(E_2-\mathcal{V})^2]\Psi_L=0
\end{equation}
for the large component, $\Psi_L=(1+\beta)\Psi/2$, of the Dirac
spinor.

Thus, we conclude that the proposed wave equation based on the
extension of the $SL(2,C)$ group has correct one-particle limits
and coincide with those obtained in other approaches to describing
the relativistic fermion-boson system.

\section{Conclusions}
\label{part5}

In the present work, the extension of the $SL(2,C)$ group to the
$Sp(4,C)$ one has been used to construct the relativistic wave
equation for the system consisted of the spin-$1/2$ and spin-$0$
particles. The obtained equation involves the potentials of both
the Lorentz-vector and Lorentz-tensor structure. With the
Lorentz-vector potentials introduced by a minimal substitution on
the four-momenta, this equation reproduces the known equation by
Kr\'{o}likowski \cite{krolikowski}, which is also derived by
reducing the Bethe-Salpeter equation \cite{tanaka}, and the
equations of the relativistic quantum mechanics with constraints
\cite{sazdjian86a,sazdjian86b,crater87}.

However, including the Lorentz-tensor potentials in our equation,
in addition to the Lorentz-vector ones, permits us to treat more
wide range of problems concerning fermion-boson interactions. It
should be pointed that the Lorentz-tensor interaction is involved
in other means than in the Breit approach \cite{kelkar}. Instead
of expanding the scattering amplitude in the $1 /c^2$-powers, that
leads to nonlocal terms in the potential \cite{kelkar}, we
introduce the local Lorentz-tensor potentials through the
non-minimal substitution on the four-momenta. In the fermion-boson
equation, the contribution of the so involved interaction
resembles the tensor coupling term appeared in the effective Dirac
Lagrangian of the relativistic mean-field theories of nuclei
\cite{furnstahl} and in the relativistic models of the Hartree
approach \cite{mao,alberto}. Thus, the constructed fermion-boson
equation is suitable for divorcing from each other the effects of
the tensor coupling and the relative motion. At last, the obtained
equation possesses the exact solutions of the harmonic-oscillator
type that will be published elsewhere.

\section*{Acknowledgments}

We thank Prof. T. Tanaka for drawing our attention to Ref.~
\cite{tanaka} and discussing its details. This research was
supported by a grant N 0106U000782 from the Ministry of Education
and Science of Ukraine which is gratefully acknowledged.


\end{document}